\documentclass[aps,prb,reprint,floatfix]{revtex4-1}
\usepackage{graphicx}%
\date{\today}

\begin{document}
\title{Niobium nitride-based normal metal-insulator-superconductor  tunnel junction microthermometer} 
\author{S. Chaudhuri}
\author{M. R. Nevala }
\author{I. J. Maasilta} \email{maasilta@jyu.fi}
\affiliation{Nanoscience Center, Department of Physics, P. O. Box 35, FI-40014 University of Jyv\"askyl\"a, Finland} 
\begin{abstract}

We have successfully fabricated  Cu-AlO$ _{x} $-Al-NbN normal metal-insulator-superconductor (NIS)  tunnel  junction devices, 
using pulsed laser deposition (PLD) for NbN film growth, and electron-beam lithography and shadow evaporation for the final device fabrication. The subgap conductance of these devices exhibit a strong temperature dependence, rendering them suitable for thermometry from $\sim$ 0.1 K all the way up to the superconducting transition temperature of the NbN layer, which was here $\sim 11$ K, but could be extended up to $\sim 16$ K in our PLD chamber. Our data fits well to the single particle  NIS tunnel junction theory, 
with an observed proximised superconducting gap value $\sim $ 1 meV for a 40 nm thick Al overlayer.  Although this high value of the  superconducting energy gap is promising  for potential electronic NIS cooling applications as well, the high value of the tunneling resistance   
  inhibits electronic cooling in the present devices. Such opaque barriers are,  however,  ideal for thermometry purposes as self-induced thermal effects are thus minimized.

\end{abstract}
\maketitle

Tunnel junction based superconducting devices have the potential to revolutionize low temperature thermometry techniques\cite{RMP}, with normal metal-insulator-superconductor (NIS) junctions most widely used so far. The operational temperature range  of NIS devices is set by the  transition temperature ($ T_C $) of the  superconductor. So far, aluminum (Al) based  superconducting devices have been successfully implemented for accurate thermometry and bolometry in the sub 1 K range \cite{Nahum,Leivo,Chouvaev,PanuPRL,JLTP,schmidt}.  A natural continuation of this line of research is the extension of thermometry to higher temperatures,  simply by using higher $ T_C $ materials as the superconducting electrode. 
  
Traditionally, niobium (Nb) has dominated the choice of  an  intermediate $ T_C $ (up to 9 K)  superconductor and  we have recently addressed the feasibility of Nb based NIS devices as thermometers and electronic coolers \cite{Nb}. Another  potentially interesting material  is niobium nitride (NbN), as the $ T_C $ of NbN thin films are typically as high as 15-16 K\cite{gavaler,oya:1389,wang2,NbNieee}, and in extreme cases even above 17 K\cite{keskar,gurvitch}, predicting a doubling of the thermometry range compared to Nb.  However, unlike for Nb based NIS devices where Nb can be  deposited in-situ using electron-beam evaporation in an ultra high vacuum (UHV) environment \cite{Nb}, the growth of NbN is not as straightforward. The $ T_C $  of NbN is an extremely sensitive function of the niobium-to-nitrogen stoichiometric ratio, and superconductivity is easily suppressed or even  destroyed  for  slight  off-stoichiometry \cite{NbNieee, oya:1389,wang2}. 

High-quality NbN based Josephson junctions have also been fabricated using several barrier materials, such as AlN \cite{wang:2034,qiu} and MgO \cite{shoji:1098,kawakami,leduc,stern}. In most cases, the superconducting electrodes and the insulating layers have been fabricated {\em in-situ} using sputter deposition techniques. 
In this paper, we report the successful fabrication of micron-scale Cu-AlO$ _{x} $-Al-NbN NIS tunnel  junctions  using a simple and straightforward process that employs pulsed laser deposition (PLD) of NbN films, electron beam lithography, reactive ion etching  and shadow angle evaporation, and {\em ex-situ} fabrication of thermally oxidized Al barriers. The temperature dependence of the current-voltage characteristics of the devices follow the simple one-particle NIS tunneling model, thus demonstrating the application of the devices in sensitive thermometry and a future potential for electronic cooling \cite{RMP,jtm}.


\begin{figure}
\includegraphics[width=0.8\columnwidth]{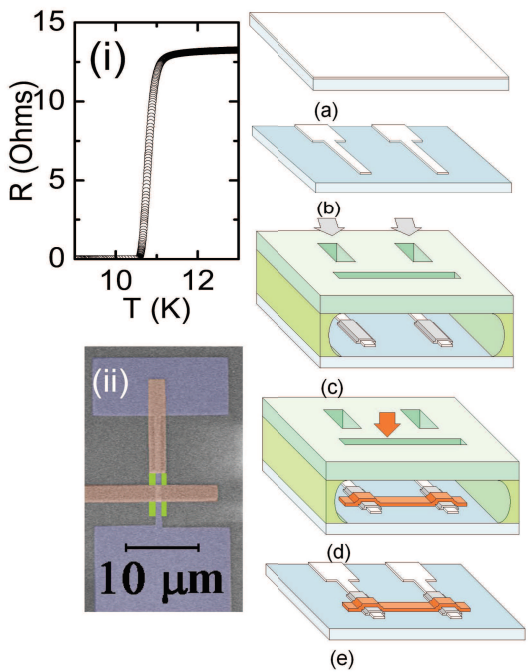}
\caption{  [Color online] (i) Temperature dependence of the resistance of the bare NbN film exhibiting superconducting transition at $\sim$ 10.8 K. 
(a)-(e) Schematics of the different steps involved  in the fabrication of a Cu-Al-AlO$ _{x} $-NbN  tunnel junction device. (a) A thin film of NbN  (white) on MgO (blue). (b) Fabrication NbN electrodes using EBL and RIE. (c) A double resist  (green layers) based lithography followed by development of resists and creation of undercuts. Angle evaporation of Al (grey)  followed by thermal oxidation. (d) Deposition of copper (red) from normal angle. (e) The final double junction device after removal of unexposed resists. (ii) False color coded scanning electron micrograph of a single NIS junction. The blue, green and orange colors represent NbN, Al and Cu respectively. The junction area is $ \sim $ 1 $ \times $ 2 $\mu m ^{2} $, while the Cu island dimensions are 38.5 $\mu m$ $\times$ 2 $\mu m$  $\times$ 80 $nm$.  }\label{Fig1}
\end{figure}
The devices were fabricated in the following steps, outlined in figure~\ref{Fig1}(a)-(e): First, a thin film  ($ \sim $  30 nm) of superconducting NbN was deposited on lattice matched single crystals of [100] oriented MgO using pulsed laser deposition technique, as described elsewhere\cite{NbNieee} (silicon substrates could also be used, if one is willing to tolerate a degradation of $T_C$). The measured $T_C$  of  the film used here  was $\sim$ 10.8 K [figure~\ref{Fig1}(i)], although $T_C$s up to $\sim 16$ K have been achieved with our setup in optimal conditions.  The  fabricated NbN film was then patterned  to make the superconducting contact pads and electrodes using electron beam lithography (EBL) and reactive ion etching (RIE) [Fig. ~\ref{Fig1}(b)].  A positive PMMA resist of thickness 400 nm served as the RIE etch mask,  while a mixture of CHF$_3$ (50 sccm) and O$_2$ (5 sccm)  was used in the RIE at a power 100 W and pressure 55 mTorr. In an second overlay EBL step, the rest of the device was patterned using a bilayer resist (described in detail in Refs. \cite{NbNieee, Nb}) to achieve a large undercut stencil mask structure suitable for angle evaporation and lift-off [Fig. ~\ref{Fig1}(c)] in an ultra-high vacuum chamber. First, an evaporation of 40 nm thick Al was performed from an angle of 20 degrees with respect to the plane of the substrate, forming two rectangular areas on top of the NbN electrodes [Fig. ~\ref{Fig1}(c)], after which the Al areas were oxidized {\em in-situ} at room temperature in 50 mbar of pure O$_2 $  for four minutes, to grow  the AlO$ _{x} $ tunnel barriers. Finally,   a 80 nm thick Cu layer was evaporated from the normal angle [Fig. ~\ref{Fig1}(d)], forming thus the final device with a Cu normal metal wire between two NbN/Al/AlOx/Cu NIS junctions [Fig. ~\ref{Fig1}(e)]. A scanning electron micrograph (SEM) of one NIS junction is shown in figure~\ref{Fig1}(ii), with a junction area of $ \sim $ 1 $ \times $ 2 $\mu m ^{2} $. 
\begin{figure}
\includegraphics[width=1\columnwidth]{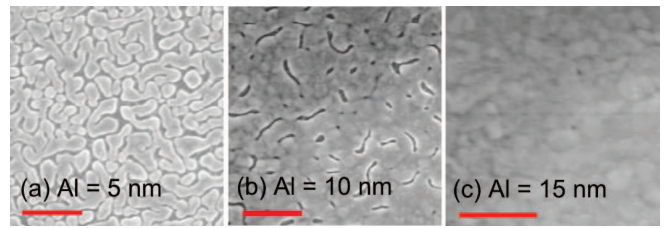}
\caption{ [Color online] Evolution of the surface morphology of the Al overlayer on NbN for various thicknesses of Al (SEM images). The morphology changes from island-type to continuous with increasing Al thickness. The horizontal scale bar at the bottom of each image is 100 nm. }\label{fig2}
\end{figure}

AlO$ _{x} $ was used here as the barrier material because of its proven excellent and controllable properties in NIS devices. However,  the downside with this choice is that the tunneling current is mostly sensitive to the density of states (DOS) in the proximized Al layer, which has a reduced gap compared to the bulk of the NbN electrode. Ideally, the Al overlayer should thus be as thin as possible, but fully covering the NbN surface and the edges. To study the wetting performance of evaporated Al on NbN, we deposited various thicknesses of Al films on top of NbN, shown   in figure~\ref{fig2}. It is seen that 5 nm thick Al exhibits island-type growth, the 10 nm thickness still has pores, but for 15 nm the coverage is mostly uniform.  For this reason, we tried to make devices with Al thickness 20 nm, but the resulting current-voltage characteristics were leaky and not ideal. One reason for that could have been poor step coverage, as the underlying NbN wire thickness was chosen to be 30 nm in order to minimize the thickness dependent suppression of $ T_C $ for ultrathin NbN films\cite{kang}. The more ideally behaving devices described below were thus fabricated using 40 nm of Al to ensure step coverage. 

\begin{figure}
\includegraphics[width=1\columnwidth]{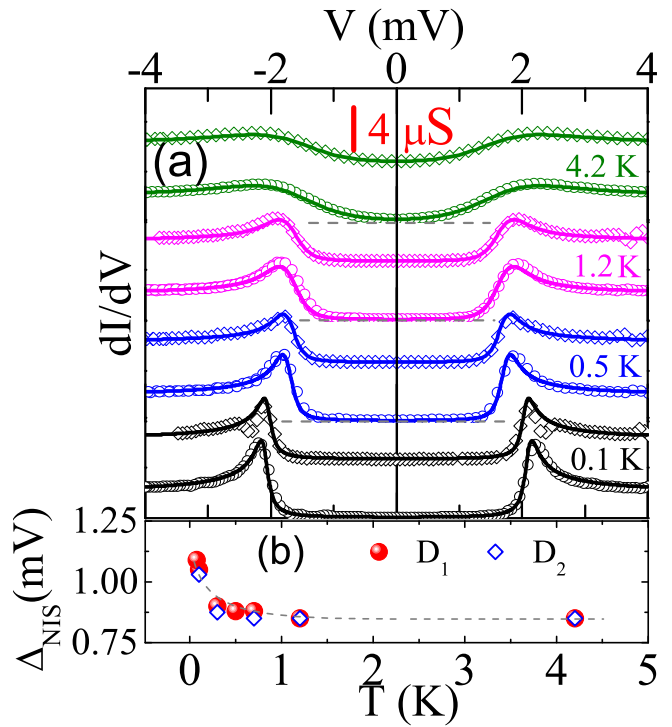}
\caption{ [Color online] (a) Differential conductance  characteristics of two  Cu-AlO$ _{x} $-Al-NbN double junction SINIS devices $D _{1} $ (circles) and $ D_{2} $ (diamonds) at various $T_{Bath}$ between 0.1 K and 4.2 K. 
The open symbols denote the measured response, while the solid lines are the corresponding theoretical fits assuming a single particle NIS model with broadened superconducting density of states \cite{RMP,PanuPRL}. For all the fits the broadening (Dynes) parameter $ \Gamma$ was kept fixed at  4 $ \times $ 10$ ^{-2} $ but  $\Delta $ was varied.  (b) The evolution of the fitted value of $\Delta$ with temperature for both devices.}\label{Fig3} 
\end{figure}
 
 The current-voltage and conductance-voltage measurements were carried out using a He$^3$-He$^4$ dilution refrigerator with a base temperature of $\sim 60$ mK.  The measurement lines had two stages of RC filters, one at 4 K and the other at base temperature, and microwave filtering  between the RC filters was achieved with the help of Thermocoax cables \cite{zorin:4296} of length 1.5 m. In figure~\ref{Fig3} we   show the  applied bias voltage  ($ V $)  dependence of the differential conductance ($ dI/dV $) at several bath temperatures ($ T_{Bath} $) for two devices  $D _{1} $ and  $ D_{2} $  fabricated from the same NbN film, measured using a lock-in amplifier \cite{note}.  The value of the total tunneling resistance $R_{T,tot}$ was $ \sim $ 630  and 770 k$\Omega $ for device $ D_1 $ and $ D_2 $, respectively, which translates to a high average  single junction specific resistance of  $ \sim $ 0.63-0.77 M$ \Omega $ $ \mu m^2 $, in agreement with previous studies of AlOx barriers on NbN \cite{talvacchio}. This can be compared with the observed values \cite{Nb,PanuPRL} $\sim 10 $k$\Omega $ $ \mu m^2 $ and $\sim 1-2 $k$\Omega $ $ \mu m^2 $ for Nb/Al/AlOx/Cu and Al/AlOx/Cu junctions, respectively, fabricated in the same chamber and with fairly similar oxidation parameters as in this work. Nevertheless, the simplest single particle NIS tunneling model \cite{RMP,Nb} fits the data, and especially the temperature dependence in the subgap region fits nearly perfectly (Fig.~\ref{Fig3}). The value of the energy gap $\Delta $ obtained from the fits evolves from  $ \sim $1.1  mV at  0.07 K to 0.85 mV at 1.2 K [Fig. ~\ref{Fig3} b)].

The fact that the simple NIS model fits our data well with $\Delta \sim 1$ mV about five times higher than the Al bulk gap $\Delta \sim 0.2$ mV indicates that the Al layer is well proximised by the NbN. Such proximity induced  enhancement  of the gap value of Al has been observed in Nb-Al-AlO$ _x $-Al-Nb junctions before \cite{Golubov,zehnder} and a microscopic theoretical  model  has been developed \cite{Golubov}. According to this model, the temperature dependence of the gap seen in tunneling experiments is no longer given by the usual BCS result, but depends on the material parameters, the thickness of the Al layer and the boundary transparency between the two superconductors. It is clearly seen that in our case the temperature dependence of the gap is not  the usual BCS type, but follows roughly the proximity effect trends shown in Refs. \cite{Golubov,zehnder}.  Perhaps the most important conclusion from the theory is that although the tunneling gap is suppressed compared to the bulk NbN gap value, it does not close until at the NbN $T_C$, permitting thus thermometry all the way up to that temperature. 
Thus by optimizing the NbN film stochiometry, it will possible to use NIS thermometry up to $\sim 16-17$ K with the device concept presented here. In addition, the large observed tunneling resistance also helps in thermometry, as the Joule heating due biasing is very small, thus keeping the thermometer more easily in equilibrium with the substrate.    

\begin{figure}
\includegraphics[width=1\columnwidth]{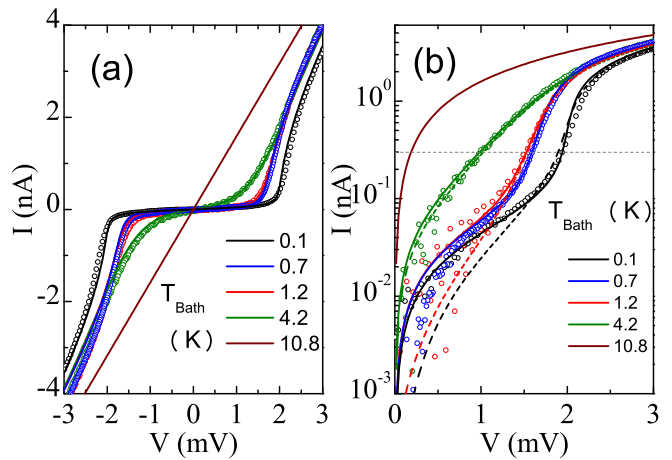}
\caption{ [Color online] Temperature dependence of  current-voltage characteristics of the device $ D_1 $ at various $T_{Bath}$ plotted in (a) linear and (b) log scale. The dots are the experimental data while the solid  lines are the corresponding theoretical fits using the Dynes model for broadening, whereas the dashed lines show some best fits using the strong-coupling model (see text).  The fitting parameters are identical to that shown in figure~\ref{Fig3} except here  $ \Gamma $ = 2.4-3$ \times$ $10^{-2}$, and $\delta=5 \times $10 $^{-2}$ for the strong coupling fits. The horizontal dashed line in (b) represents a constant current bias of 0.3 nA.}\label{Fig5} 
\end{figure}

In figure~\ref{Fig5} the current-voltage ($I$-$V$) characteristics at various $T_{Bath}$ are shown for device $ D_1 $ in (a) linear and (b) log scale, respectively, together with the corresponding theoretical fits based on the single-particle tunneling model  $I=\frac{1}{eR_{T,tot}}\int_{-\infty}^{\infty}\!\! d\epsilon N_{S}(\epsilon)[f_{N}(\epsilon-eV/2)-f_{N}(\epsilon+eV/2)]$,  where $f_{N}(\epsilon)$ is the Fermi function in the Cu wire, and $N_{S}(\epsilon )$ is the normalized broadened superconducting quasiparticle DOS in the Dynes model \cite{Dynes}  $N_{S}(\epsilon, T_{S})  =\left | {\rm Re} \left ( \frac{\epsilon+i\Gamma}{\sqrt{(\epsilon+i\Gamma)^{2}-\Delta^{2}}} \right ) \right |$, and a constant electron temperature $T_e = T_{Bath}$ was used. We see that for a constant current bias  the voltage drop across the device changes with $T_{Bath}$ as expected from the simple theory, and  thus the typical thermometric responsivity of a NIS junction \cite{RMP} exists from 0.1 K to all the way up to 10.8 K ($  T_C$). 
The theoretical fits are very good, obtained with $ \Gamma/\Delta $ = 2.4-3$ \times $10$ ^{-2} $, consistent with the $dI/dV$ data if the effect of ac-excitation smearing is taken into account. This value of $\Gamma/\Delta $ aprroximately two times smaller than what we have observed for Nb-based NIS junctions\cite{Nb}, but still orders of magnitude larger than for Al/AlOx/Cu NIS junctions \cite{JLTP,PekolaDO}, and in agreement with data from large area NbN junctions \cite{PhysRevB.79.094509}. We should point out that these and other observed values of the DOS broadening should not necessarily be taken as materials parameters, as the external fluctuations from the electromagnetic environment can influence the broadening strongly \cite{PekolaDO,saira}. For strong coupling superconductors such as NbN, however, one would expect theoretically \cite{Wolf,Mitrovic} that the gap $\Delta$ is a complex number such that without environmental effects $N_{S}(\epsilon)  =\left | {\rm Re} \left ( \frac{\epsilon}{\sqrt{\epsilon^{2}-(\Delta+i\delta)^{2}}} \right ) \right |$. Good fits to subgap current in Nb SIS junctions have recently been obtained with such a model \cite{Noguchi}.  
We also tried alternative strong-coupling theory fits to the data, shown as dotted lines in figure~\ref{Fig5} (b), but with far worse agreement with the data in the deep subgap as compared to the Dynes model fits.  

\begin{figure}
\includegraphics[width=1\columnwidth]{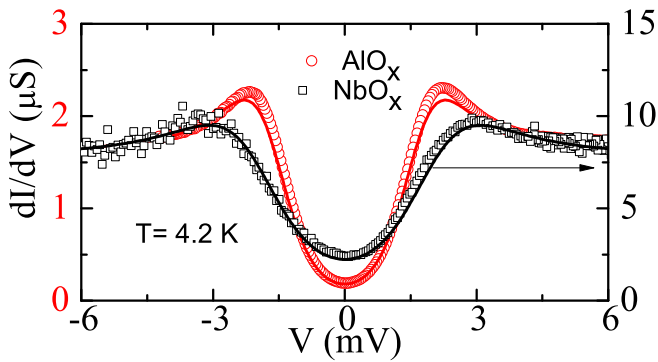}
\caption{ [Color online] Measured conductance  characteristics of a  directly oxidized Cu-NbO$ _{x} $-NbN device (squares) compared to device $D _{1} $ (circles) at 4.2 K,   with the theoretical fits to Dynes broadened theory indicated by the solid lines. The directly oxidized device has a large $\Gamma/\Delta$ =0.2  }\label{Fig4} 
\end{figure}

Successful NbN tunnel junctions have also been made by oxidizing the NbN surface directly, in room atmosphere at room temperature \cite{dover:764} or in oxygen atmosphere at 200 $\deg$ C \cite{PhysRevB.79.094509}. We also tried a variation of direct oxidation, by  sputter cleaning the NbN surface in-situ, then oxidising the NbN at 350 mbar of oxygen pressure  at room temperature for 45 minutes before depositing  the copper. Although barrier formation was successful, the process was not very reproducible, and the obtained subgap conductance characteristics (Fig. ~\ref{Fig4}) were more broadened than with the Al overlayer method, rendering the direct oxidation process less useful. However, the specific resistances of the directly oxidized junctions were clearly smaller ( $ \sim $ 40 k$ \Omega $ $ \mu m^2 $) than the ones with AlO$ _x $ barriers. Further studies with oxidation at elevated temperatures should therefore be tried.

In addition to applications in thermometry, high quality NIS junctions can also be used as electronic microrefrigerators \cite{jtm,RMP}. Good performance has only been obtained with Al as the superconductor \cite{jtm,oneil}, but the low value of the superconducting gap in Al limits the useful cooling regime to 0.3 K and below.  We have also recently tried Nb as well \cite{Nb} with some limited success, but theoretically NbN junctions could offer better performance at even higher temperatures. The junctions fabricated here, however, had much too high specific resistance for cooler applications (cooling power is inversely proportional to $R_{T}$). Nevertheless, the observed subgap current level,$ \Gamma/\Delta $ = 2-3$ \times $10$ ^{-2} $, does not prevent cooling altogether, but only limits the operational temperature range. A smaller $\Gamma$ would lead to improved cooling regime and power.

In conclusion, we have successfully fabricated micron-scale Cu-AlO$ _{x} $-Al-NbN  NIS tunnel  junction devices using pulsed laser deposited NbN films and e-beam lithography.  The  temperature dependent tunneling characteristics were ideal enough for wide temperature range sensitive low-temperature thermometry from 0.1 K up to the critical temperature of NbN, which for PLD films can be as high as 16-17 K. The combined high responsivity and modest operational temperature requirements may offer advantages in some sensitive bolometry applications, as well.   
 
This research was supported by Academy of Finland project numbers 128532 and 260880, useful discussions with J. Ullom are acknowledged.   

%


\begin{thebibliography}{36}%
\makeatletter
\providecommand \@ifxundefined [1]{%
 \@ifx{#1\undefined}
}%
\providecommand \@ifnum [1]{%
 \ifnum #1\expandafter \@firstoftwo
 \else \expandafter \@secondoftwo
 \fi
}%
\providecommand \@ifx [1]{%
 \ifx #1\expandafter \@firstoftwo
 \else \expandafter \@secondoftwo
 \fi
}%
\providecommand \natexlab [1]{#1}%
\providecommand \enquote  [1]{``#1''}%
\providecommand \bibnamefont  [1]{#1}%
\providecommand \bibfnamefont [1]{#1}%
\providecommand \citenamefont [1]{#1}%
\providecommand \href@noop [0]{\@secondoftwo}%
\providecommand \href [0]{\begingroup \@sanitize@url \@href}%
\providecommand \@href[1]{\@@startlink{#1}\@@href}%
\providecommand \@@href[1]{\endgroup#1\@@endlink}%
\providecommand \@sanitize@url [0]{\catcode `\\12\catcode `\$12\catcode
  `\&12\catcode `\#12\catcode `\^12\catcode `\_12\catcode `\%12\relax}%
\providecommand \@@startlink[1]{}%
\providecommand \@@endlink[0]{}%
\providecommand \url  [0]{\begingroup\@sanitize@url \@url }%
\providecommand \@url [1]{\endgroup\@href {#1}{\urlprefix }}%
\providecommand \urlprefix  [0]{URL }%
\providecommand \Eprint [0]{\href }%
\providecommand \doibase [0]{http://dx.doi.org/}%
\providecommand \selectlanguage [0]{\@gobble}%
\providecommand \bibinfo  [0]{\@secondoftwo}%
\providecommand \bibfield  [0]{\@secondoftwo}%
\providecommand \translation [1]{[#1]}%
\providecommand \BibitemOpen [0]{}%
\providecommand \bibitemStop [0]{}%
\providecommand \bibitemNoStop [0]{.\EOS\space}%
\providecommand \EOS [0]{\spacefactor3000\relax}%
\providecommand \BibitemShut  [1]{\csname bibitem#1\endcsname}%
\let\auto@bib@innerbib\@empty
\bibitem [{\citenamefont {Giazotto}\ \emph {et~al.}(2006)\citenamefont
  {Giazotto}, \citenamefont {Heikkil\"a}, \citenamefont {Luukanen},
  \citenamefont {Savin},\ and\ \citenamefont {Pekola}}]{RMP}%
  \BibitemOpen
  \bibfield  {author} {\bibinfo {author} {\bibfnamefont {F.}~\bibnamefont
  {Giazotto}}, \bibinfo {author} {\bibfnamefont {T.~T.}\ \bibnamefont
  {Heikkil\"a}}, \bibinfo {author} {\bibfnamefont {A.}~\bibnamefont
  {Luukanen}}, \bibinfo {author} {\bibfnamefont {A.~M.}\ \bibnamefont {Savin}},
  \ and\ \bibinfo {author} {\bibfnamefont {J.~P.}\ \bibnamefont {Pekola}},\
  }\href {\doibase 10.1103/RevModPhys.78.217} {\bibfield  {journal} {\bibinfo
  {journal} {Rev. Mod. Phys.}\ }\textbf {\bibinfo {volume} {78}},\ \bibinfo
  {pages} {217} (\bibinfo {year} {2006})}\BibitemShut {NoStop}%
\bibitem [{\citenamefont {Nahum}\ and\ \citenamefont {Martinis}(1993)}]{Nahum}%
  \BibitemOpen
  \bibfield  {author} {\bibinfo {author} {\bibfnamefont {M.}~\bibnamefont
  {Nahum}}\ and\ \bibinfo {author} {\bibfnamefont {J.~M.}\ \bibnamefont
  {Martinis}},\ }\href {\doibase 10.1063/1.110237} {\bibfield  {journal}
  {\bibinfo  {journal} {Appl. Phys. Lett.}\ }\textbf {\bibinfo {volume} {63}},\
  \bibinfo {pages} {3075} (\bibinfo {year} {1993})}\BibitemShut {NoStop}%
\bibitem [{\citenamefont {Leivo}\ \emph {et~al.}(1996)\citenamefont {Leivo},
  \citenamefont {Pekola},\ and\ \citenamefont {Averin}}]{Leivo}%
  \BibitemOpen
  \bibfield  {author} {\bibinfo {author} {\bibfnamefont {M.~M.}\ \bibnamefont
  {Leivo}}, \bibinfo {author} {\bibfnamefont {J.~P.}\ \bibnamefont {Pekola}}, \
  and\ \bibinfo {author} {\bibfnamefont {D.}~\bibnamefont {Averin}},\
  }\href@noop {} {\bibfield  {journal} {\bibinfo  {journal} {Appl. Phys.
  Lett.}\ }\textbf {\bibinfo {volume} {68}},\ \bibinfo {pages} {1996} (\bibinfo
  {year} {1996})}\BibitemShut {NoStop}%
\bibitem [{\citenamefont {Chouvaev}\ \emph {et~al.}(1999)\citenamefont
  {Chouvaev}, \citenamefont {Kuzmin},\ and\ \citenamefont
  {Tarasov}}]{Chouvaev}%
  \BibitemOpen
  \bibfield  {author} {\bibinfo {author} {\bibfnamefont {D.}~\bibnamefont
  {Chouvaev}}, \bibinfo {author} {\bibfnamefont {L.}~\bibnamefont {Kuzmin}}, \
  and\ \bibinfo {author} {\bibfnamefont {M.}~\bibnamefont {Tarasov}},\
  }\href@noop {} {\bibfield  {journal} {\bibinfo  {journal} {Supercond. Sci.
  Technol.}\ }\textbf {\bibinfo {volume} {12}},\ \bibinfo {pages} {985}
  (\bibinfo {year} {1999})}\BibitemShut {NoStop}%
\bibitem [{\citenamefont {Koppinen}\ and\ \citenamefont
  {Maasilta}(2009)}]{PanuPRL}%
  \BibitemOpen
  \bibfield  {author} {\bibinfo {author} {\bibfnamefont {P.~J.}\ \bibnamefont
  {Koppinen}}\ and\ \bibinfo {author} {\bibfnamefont {I.~J.}\ \bibnamefont
  {Maasilta}},\ }\href {\doibase 10.1103/PhysRevLett.102.165502} {\bibfield
  {journal} {\bibinfo  {journal} {Phys. Rev. Lett.}\ }\textbf {\bibinfo
  {volume} {102}},\ \bibinfo {pages} {165502} (\bibinfo {year}
  {2009})}\BibitemShut {NoStop}%
\bibitem [{\citenamefont {Koppinen}\ \emph {et~al.}(2009)\citenamefont
  {Koppinen}, \citenamefont {K\"uhn},\ and\ \citenamefont {Maasilta}}]{JLTP}%
  \BibitemOpen
  \bibfield  {author} {\bibinfo {author} {\bibfnamefont {P.~J.}\ \bibnamefont
  {Koppinen}}, \bibinfo {author} {\bibfnamefont {T.}~\bibnamefont {K\"uhn}}, \
  and\ \bibinfo {author} {\bibfnamefont {I.~J.}\ \bibnamefont {Maasilta}},\
  }\href@noop {} {\bibfield  {journal} {\bibinfo  {journal} {J. Low Temp.
  Phys.}\ }\textbf {\bibinfo {volume} {154}},\ \bibinfo {pages} {179} (\bibinfo
  {year} {2009})}\BibitemShut {NoStop}%
\bibitem [{\citenamefont {Schmidt}\ \emph {et~al.}(2005)\citenamefont
  {Schmidt}, \citenamefont {Lehnert}, \citenamefont {Clark}, \citenamefont
  {Duncan}, \citenamefont {Irwin}, \citenamefont {Miller},\ and\ \citenamefont
  {Ullom}}]{schmidt}%
  \BibitemOpen
  \bibfield  {author} {\bibinfo {author} {\bibfnamefont {D.~R.}\ \bibnamefont
  {Schmidt}}, \bibinfo {author} {\bibfnamefont {K.~W.}\ \bibnamefont
  {Lehnert}}, \bibinfo {author} {\bibfnamefont {A.~M.}\ \bibnamefont {Clark}},
  \bibinfo {author} {\bibfnamefont {W.~D.}\ \bibnamefont {Duncan}}, \bibinfo
  {author} {\bibfnamefont {K.~D.}\ \bibnamefont {Irwin}}, \bibinfo {author}
  {\bibfnamefont {N.}~\bibnamefont {Miller}}, \ and\ \bibinfo {author}
  {\bibfnamefont {J.~N.}\ \bibnamefont {Ullom}},\ }\href@noop {} {\bibfield
  {journal} {\bibinfo  {journal} {Appl. Phys. Lett.}\ }\textbf {\bibinfo
  {volume} {86}},\ \bibinfo {pages} {053505} (\bibinfo {year}
  {2005})}\BibitemShut {NoStop}%
\bibitem [{\citenamefont {Nevala}\ \emph {et~al.}(2012)\citenamefont {Nevala},
  \citenamefont {Chaudhuri}, \citenamefont {Halkosaari}, \citenamefont
  {Karvonen},\ and\ \citenamefont {Maasilta}}]{Nb}%
  \BibitemOpen
  \bibfield  {author} {\bibinfo {author} {\bibfnamefont {M.~R.}\ \bibnamefont
  {Nevala}}, \bibinfo {author} {\bibfnamefont {S.}~\bibnamefont {Chaudhuri}},
  \bibinfo {author} {\bibfnamefont {J.}~\bibnamefont {Halkosaari}}, \bibinfo
  {author} {\bibfnamefont {J.~T.}\ \bibnamefont {Karvonen}}, \ and\ \bibinfo
  {author} {\bibfnamefont {I.~J.}\ \bibnamefont {Maasilta}},\ }\href {\doibase
  10.1063/1.4751355} {\bibfield  {journal} {\bibinfo  {journal} {Appl. Phys.
  Lett.}\ }\textbf {\bibinfo {volume} {101}},\ \bibinfo {eid} {112601}
  (\bibinfo {year} {2012})}\BibitemShut {NoStop}%
\bibitem [{\citenamefont {Gavaler}\ \emph {et~al.}(1969)\citenamefont
  {Gavaler}, \citenamefont {Hulm}, \citenamefont {Janocko},\ and\ \citenamefont
  {JOnes}}]{gavaler}%
  \BibitemOpen
  \bibfield  {author} {\bibinfo {author} {\bibfnamefont {J.~R.}\ \bibnamefont
  {Gavaler}}, \bibinfo {author} {\bibfnamefont {J.~K.}\ \bibnamefont {Hulm}},
  \bibinfo {author} {\bibfnamefont {M.~A.}\ \bibnamefont {Janocko}}, \ and\
  \bibinfo {author} {\bibfnamefont {C.~K.}\ \bibnamefont {JOnes}},\ }\href@noop
  {} {\bibfield  {journal} {\bibinfo  {journal} {J. Vac. Sci. Technol.}\
  }\textbf {\bibinfo {volume} {6}},\ \bibinfo {pages} {177} (\bibinfo {year}
  {1969})}\BibitemShut {NoStop}%
\bibitem [{\citenamefont {i.~Oya}\ and\ \citenamefont
  {Onodera}(1974)}]{oya:1389}%
  \BibitemOpen
  \bibfield  {author} {\bibinfo {author} {\bibfnamefont {G.}~\bibnamefont
  {i.~Oya}}\ and\ \bibinfo {author} {\bibfnamefont {Y.}~\bibnamefont
  {Onodera}},\ }\href {\doibase 10.1063/1.1663418} {\bibfield  {journal}
  {\bibinfo  {journal} {J. Appl. Phys.}\ }\textbf {\bibinfo {volume} {45}},\
  \bibinfo {pages} {1389} (\bibinfo {year} {1974})}\BibitemShut {NoStop}%
\bibitem [{\citenamefont {Wang}\ \emph {et~al.}(1996)\citenamefont {Wang},
  \citenamefont {Kawakami}, \citenamefont {Uzawa},\ and\ \citenamefont
  {Komiyama}}]{wang2}%
  \BibitemOpen
  \bibfield  {author} {\bibinfo {author} {\bibfnamefont {Z.}~\bibnamefont
  {Wang}}, \bibinfo {author} {\bibfnamefont {A.}~\bibnamefont {Kawakami}},
  \bibinfo {author} {\bibfnamefont {Y.}~\bibnamefont {Uzawa}}, \ and\ \bibinfo
  {author} {\bibfnamefont {B.}~\bibnamefont {Komiyama}},\ }\href@noop {}
  {\bibfield  {journal} {\bibinfo  {journal} {J. Appl. Phys.}\ }\textbf
  {\bibinfo {volume} {79}},\ \bibinfo {pages} {7837} (\bibinfo {year}
  {1996})}\BibitemShut {NoStop}%
\bibitem [{\citenamefont {Chaudhuri}\ \emph {et~al.}(2011)\citenamefont
  {Chaudhuri}, \citenamefont {Nevala}, \citenamefont {Hakkarainen},
  \citenamefont {Niemi},\ and\ \citenamefont {Maasilta}}]{NbNieee}%
  \BibitemOpen
  \bibfield  {author} {\bibinfo {author} {\bibfnamefont {S.}~\bibnamefont
  {Chaudhuri}}, \bibinfo {author} {\bibfnamefont {M.}~\bibnamefont {Nevala}},
  \bibinfo {author} {\bibfnamefont {T.}~\bibnamefont {Hakkarainen}}, \bibinfo
  {author} {\bibfnamefont {T.}~\bibnamefont {Niemi}}, \ and\ \bibinfo {author}
  {\bibfnamefont {I.}~\bibnamefont {Maasilta}},\ }\href {\doibase
  10.1109/TASC.2010.2081656} {\bibfield  {journal} {\bibinfo  {journal} {IEEE
  Trans. Appl. Supercond.}\ }\textbf {\bibinfo {volume} {21}},\ \bibinfo
  {pages} {143 } (\bibinfo {year} {2011})}\BibitemShut {NoStop}%
\bibitem [{\citenamefont {Keskar}\ \emph {et~al.}(1971)\citenamefont {Keskar},
  \citenamefont {Yamashita},\ and\ \citenamefont {Onodera}}]{keskar}%
  \BibitemOpen
  \bibfield  {author} {\bibinfo {author} {\bibfnamefont {K.~S.}\ \bibnamefont
  {Keskar}}, \bibinfo {author} {\bibfnamefont {T.}~\bibnamefont {Yamashita}}, \
  and\ \bibinfo {author} {\bibfnamefont {Y.}~\bibnamefont {Onodera}},\
  }\href@noop {} {\bibfield  {journal} {\bibinfo  {journal} {Jap. J. Appl.
  Phys.}\ }\textbf {\bibinfo {volume} {10}},\ \bibinfo {pages} {370} (\bibinfo
  {year} {1971})}\BibitemShut {NoStop}%
\bibitem [{\citenamefont {Gurvitch}\ \emph {et~al.}(1985)\citenamefont
  {Gurvitch}, \citenamefont {Remeika}, \citenamefont {Rowell}, \citenamefont
  {Geerk},\ and\ \citenamefont {Lowe}}]{gurvitch}%
  \BibitemOpen
  \bibfield  {author} {\bibinfo {author} {\bibfnamefont {M.}~\bibnamefont
  {Gurvitch}}, \bibinfo {author} {\bibfnamefont {J.~P.}\ \bibnamefont
  {Remeika}}, \bibinfo {author} {\bibfnamefont {J.~M.}\ \bibnamefont {Rowell}},
  \bibinfo {author} {\bibfnamefont {J.}~\bibnamefont {Geerk}}, \ and\ \bibinfo
  {author} {\bibfnamefont {W.~P.}\ \bibnamefont {Lowe}},\ }\href@noop {}
  {\bibfield  {journal} {\bibinfo  {journal} {IEEE Trans. Mag.}\ }\textbf
  {\bibinfo {volume} {MAG-21}},\ \bibinfo {pages} {509} (\bibinfo {year}
  {1985})}\BibitemShut {NoStop}%
\bibitem [{\citenamefont {Wang}\ \emph {et~al.}(1994)\citenamefont {Wang},
  \citenamefont {Kawakami}, \citenamefont {Uzawa},\ and\ \citenamefont
  {Komiyama}}]{wang:2034}%
  \BibitemOpen
  \bibfield  {author} {\bibinfo {author} {\bibfnamefont {Z.}~\bibnamefont
  {Wang}}, \bibinfo {author} {\bibfnamefont {A.}~\bibnamefont {Kawakami}},
  \bibinfo {author} {\bibfnamefont {Y.}~\bibnamefont {Uzawa}}, \ and\ \bibinfo
  {author} {\bibfnamefont {B.}~\bibnamefont {Komiyama}},\ }\href {\doibase
  10.1063/1.111730} {\bibfield  {journal} {\bibinfo  {journal} {Appl. Phys.
  Lett.}\ }\textbf {\bibinfo {volume} {64}},\ \bibinfo {pages} {2034} (\bibinfo
  {year} {1994})}\BibitemShut {NoStop}%
\bibitem [{\citenamefont {Qiu}\ \emph {et~al.}(2011)\citenamefont {Qiu},
  \citenamefont {Terai},\ and\ \citenamefont {Wang}}]{qiu}%
  \BibitemOpen
  \bibfield  {author} {\bibinfo {author} {\bibfnamefont {W.}~\bibnamefont
  {Qiu}}, \bibinfo {author} {\bibfnamefont {H.}~\bibnamefont {Terai}}, \ and\
  \bibinfo {author} {\bibfnamefont {Z.}~\bibnamefont {Wang}},\ }\href {\doibase
  10.1109/TASC.2010.2081653} {\bibfield  {journal} {\bibinfo  {journal} {IEEE
  Trans. Appl. Supercond.}\ }\textbf {\bibinfo {volume} {21}},\ \bibinfo
  {pages} {135 } (\bibinfo {year} {2011})}\BibitemShut {NoStop}%
\bibitem [{\citenamefont {Shoji}\ \emph {et~al.}(1985)\citenamefont {Shoji},
  \citenamefont {Aoyagi}, \citenamefont {Kosaka}, \citenamefont {Shinoki},\
  and\ \citenamefont {Hayakawa}}]{shoji:1098}%
  \BibitemOpen
  \bibfield  {author} {\bibinfo {author} {\bibfnamefont {A.}~\bibnamefont
  {Shoji}}, \bibinfo {author} {\bibfnamefont {M.}~\bibnamefont {Aoyagi}},
  \bibinfo {author} {\bibfnamefont {S.}~\bibnamefont {Kosaka}}, \bibinfo
  {author} {\bibfnamefont {F.}~\bibnamefont {Shinoki}}, \ and\ \bibinfo
  {author} {\bibfnamefont {H.}~\bibnamefont {Hayakawa}},\ }\href {\doibase
  10.1063/1.95774} {\bibfield  {journal} {\bibinfo  {journal} {Appl. Phys.
  Lett.}\ }\textbf {\bibinfo {volume} {46}},\ \bibinfo {pages} {1098} (\bibinfo
  {year} {1985})}\BibitemShut {NoStop}%
\bibitem [{\citenamefont {Kawakami}\ \emph {et~al.}(2001)\citenamefont
  {Kawakami}, \citenamefont {Wang},\ and\ \citenamefont {Miki}}]{kawakami}%
  \BibitemOpen
  \bibfield  {author} {\bibinfo {author} {\bibfnamefont {A.}~\bibnamefont
  {Kawakami}}, \bibinfo {author} {\bibfnamefont {Z.}~\bibnamefont {Wang}}, \
  and\ \bibinfo {author} {\bibfnamefont {S.}~\bibnamefont {Miki}},\ }\href
  {\doibase 10.1063/1.1409583} {\bibfield  {journal} {\bibinfo  {journal} {J.
  Appl. Phys.}\ }\textbf {\bibinfo {volume} {90}},\ \bibinfo {pages} {4796}
  (\bibinfo {year} {2001})}\BibitemShut {NoStop}%
\bibitem [{\citenamefont {LeDuc}\ \emph {et~al.}(1991)\citenamefont {LeDuc},
  \citenamefont {Judas}, \citenamefont {Cypher}, \citenamefont {Bumble},
  \citenamefont {Hunt},\ and\ \citenamefont {Stern}}]{leduc}%
  \BibitemOpen
  \bibfield  {author} {\bibinfo {author} {\bibfnamefont {H.}~\bibnamefont
  {LeDuc}}, \bibinfo {author} {\bibfnamefont {A.}~\bibnamefont {Judas}},
  \bibinfo {author} {\bibfnamefont {S.}~\bibnamefont {Cypher}}, \bibinfo
  {author} {\bibfnamefont {B.}~\bibnamefont {Bumble}}, \bibinfo {author}
  {\bibfnamefont {B.}~\bibnamefont {Hunt}}, \ and\ \bibinfo {author}
  {\bibfnamefont {J.}~\bibnamefont {Stern}},\ }\href {\doibase
  10.1109/20.133890} {\bibfield  {journal} {\bibinfo  {journal} {IEEE Trans.
  Magn.}\ }\textbf {\bibinfo {volume} {27}},\ \bibinfo {pages} {3192 }
  (\bibinfo {year} {1991})}\BibitemShut {NoStop}%
\bibitem [{\citenamefont {Stern}\ \emph {et~al.}(1989)\citenamefont {Stern},
  \citenamefont {Hunt}, \citenamefont {LeDuc}, \citenamefont {Judas},
  \citenamefont {McGrath}, \citenamefont {Cypher},\ and\ \citenamefont
  {Khanna}}]{stern}%
  \BibitemOpen
  \bibfield  {author} {\bibinfo {author} {\bibfnamefont {J.}~\bibnamefont
  {Stern}}, \bibinfo {author} {\bibfnamefont {B.}~\bibnamefont {Hunt}},
  \bibinfo {author} {\bibfnamefont {H.}~\bibnamefont {LeDuc}}, \bibinfo
  {author} {\bibfnamefont {A.}~\bibnamefont {Judas}}, \bibinfo {author}
  {\bibfnamefont {W.}~\bibnamefont {McGrath}}, \bibinfo {author} {\bibfnamefont
  {S.}~\bibnamefont {Cypher}}, \ and\ \bibinfo {author} {\bibfnamefont
  {S.}~\bibnamefont {Khanna}},\ }\href {\doibase 10.1109/20.92470} {\bibfield
  {journal} {\bibinfo  {journal} {IEEE Trans. Magn.}\ }\textbf {\bibinfo
  {volume} {25}},\ \bibinfo {pages} {1054 } (\bibinfo {year}
  {1989})}\BibitemShut {NoStop}%
\bibitem [{\citenamefont {Muhonen}\ \emph {et~al.}(2012)\citenamefont
  {Muhonen}, \citenamefont {Meschke},\ and\ \citenamefont {Pekola}}]{jtm}%
  \BibitemOpen
  \bibfield  {author} {\bibinfo {author} {\bibfnamefont {J.~T.}\ \bibnamefont
  {Muhonen}}, \bibinfo {author} {\bibfnamefont {M.}~\bibnamefont {Meschke}}, \
  and\ \bibinfo {author} {\bibfnamefont {J.~P.}\ \bibnamefont {Pekola}},\
  }\href {http://stacks.iop.org/0034-4885/75/i=4/a=046501} {\bibfield
  {journal} {\bibinfo  {journal} {Rep. Prog. Phys.}\ }\textbf {\bibinfo
  {volume} {75}},\ \bibinfo {pages} {046501} (\bibinfo {year}
  {2012})}\BibitemShut {NoStop}%
\bibitem [{\citenamefont {Kang}\ \emph {et~al.}(2011)\citenamefont {Kang},
  \citenamefont {Jin}, \citenamefont {Liu}, \citenamefont {Jia}, \citenamefont
  {Chen}, \citenamefont {Ji}, \citenamefont {Xu}, \citenamefont {Wu},
  \citenamefont {Mi}, \citenamefont {Pimenov}, \citenamefont {Wu},\ and\
  \citenamefont {Wang}}]{kang}%
  \BibitemOpen
  \bibfield  {author} {\bibinfo {author} {\bibfnamefont {L.}~\bibnamefont
  {Kang}}, \bibinfo {author} {\bibfnamefont {B.~B.}\ \bibnamefont {Jin}},
  \bibinfo {author} {\bibfnamefont {X.~Y.}\ \bibnamefont {Liu}}, \bibinfo
  {author} {\bibfnamefont {X.~Q.}\ \bibnamefont {Jia}}, \bibinfo {author}
  {\bibfnamefont {J.}~\bibnamefont {Chen}}, \bibinfo {author} {\bibfnamefont
  {Z.~M.}\ \bibnamefont {Ji}}, \bibinfo {author} {\bibfnamefont {W.~W.}\
  \bibnamefont {Xu}}, \bibinfo {author} {\bibfnamefont {P.~H.}\ \bibnamefont
  {Wu}}, \bibinfo {author} {\bibfnamefont {S.~B.}\ \bibnamefont {Mi}}, \bibinfo
  {author} {\bibfnamefont {A.}~\bibnamefont {Pimenov}}, \bibinfo {author}
  {\bibfnamefont {Y.~J.}\ \bibnamefont {Wu}}, \ and\ \bibinfo {author}
  {\bibfnamefont {B.~G.}\ \bibnamefont {Wang}},\ }\href {\doibase
  10.1063/1.3518037} {\bibfield  {journal} {\bibinfo  {journal} {J. Appl.
  Phys.}\ }\textbf {\bibinfo {volume} {109}},\ \bibinfo {eid} {033908}
  (\bibinfo {year} {2011})}\BibitemShut {NoStop}%
\bibitem [{\citenamefont {Zorin}(1995)}]{zorin:4296}%
  \BibitemOpen
  \bibfield  {author} {\bibinfo {author} {\bibfnamefont {A.~B.}\ \bibnamefont
  {Zorin}},\ }\href {\doibase 10.1063/1.1145385} {\bibfield  {journal}
  {\bibinfo  {journal} {Rev. Sci. Instrum.}\ }\textbf {\bibinfo {volume}
  {66}},\ \bibinfo {pages} {4296} (\bibinfo {year} {1995})}\BibitemShut
  {NoStop}%
\bibitem [{not()}]{note}%
  \BibitemOpen
  \href@noop {} {\bibinfo  {journal} {We used an excitation voltage 0.08 mV and
  frequency of 17 Hz in the lock-in measurement.}\ }\BibitemShut {NoStop}%
\bibitem [{\citenamefont {Talvacchio}\ \emph {et~al.}(1985)\citenamefont
  {Talvacchio}, \citenamefont {Gavaler}, \citenamefont {Braginski},\ and\
  \citenamefont {Janocko}}]{talvacchio}%
  \BibitemOpen
\bibfield  {journal} {  }\bibfield  {author} {\bibinfo {author} {\bibfnamefont
  {J.}~\bibnamefont {Talvacchio}}, \bibinfo {author} {\bibfnamefont {J.~R.}\
  \bibnamefont {Gavaler}}, \bibinfo {author} {\bibfnamefont {A.~I.}\
  \bibnamefont {Braginski}}, \ and\ \bibinfo {author} {\bibfnamefont {M.~A.}\
  \bibnamefont {Janocko}},\ }\href {\doibase 10.1063/1.336234} {\bibfield
  {journal} {\bibinfo  {journal} {J. Appl. Phys.}\ }\textbf {\bibinfo {volume}
  {58}},\ \bibinfo {pages} {4638} (\bibinfo {year} {1985})}\BibitemShut
  {NoStop}%
\bibitem [{\citenamefont {Golubov}\ \emph {et~al.}(1995)\citenamefont
  {Golubov}, \citenamefont {Houwman}, \citenamefont {Gijsbertsen},
  \citenamefont {Krasnov}, \citenamefont {Flokstra}, \citenamefont {Rogalla},\
  and\ \citenamefont {Kupriyanov}}]{Golubov}%
  \BibitemOpen
  \bibfield  {author} {\bibinfo {author} {\bibfnamefont {A.~A.}\ \bibnamefont
  {Golubov}}, \bibinfo {author} {\bibfnamefont {E.~P.}\ \bibnamefont
  {Houwman}}, \bibinfo {author} {\bibfnamefont {J.~G.}\ \bibnamefont
  {Gijsbertsen}}, \bibinfo {author} {\bibfnamefont {V.~M.}\ \bibnamefont
  {Krasnov}}, \bibinfo {author} {\bibfnamefont {J.}~\bibnamefont {Flokstra}},
  \bibinfo {author} {\bibfnamefont {H.}~\bibnamefont {Rogalla}}, \ and\
  \bibinfo {author} {\bibfnamefont {M.~Y.}\ \bibnamefont {Kupriyanov}},\ }\href
  {\doibase 10.1103/PhysRevB.51.1073} {\bibfield  {journal} {\bibinfo
  {journal} {Phys. Rev. B}\ }\textbf {\bibinfo {volume} {51}},\ \bibinfo
  {pages} {1073} (\bibinfo {year} {1995})}\BibitemShut {NoStop}%
\bibitem [{\citenamefont {Zehnder}\ \emph {et~al.}(1999)\citenamefont
  {Zehnder}, \citenamefont {Lerch}, \citenamefont {Zhao}, \citenamefont
  {Nussbaumer}, \citenamefont {Kirk},\ and\ \citenamefont {Ott}}]{zehnder}%
  \BibitemOpen
  \bibfield  {author} {\bibinfo {author} {\bibfnamefont {A.}~\bibnamefont
  {Zehnder}}, \bibinfo {author} {\bibfnamefont {P.}~\bibnamefont {Lerch}},
  \bibinfo {author} {\bibfnamefont {S.~P.}\ \bibnamefont {Zhao}}, \bibinfo
  {author} {\bibfnamefont {T.}~\bibnamefont {Nussbaumer}}, \bibinfo {author}
  {\bibfnamefont {E.~C.}\ \bibnamefont {Kirk}}, \ and\ \bibinfo {author}
  {\bibfnamefont {H.~R.}\ \bibnamefont {Ott}},\ }\href {\doibase
  10.1103/PhysRevB.59.8875} {\bibfield  {journal} {\bibinfo  {journal} {Phys.
  Rev. B}\ }\textbf {\bibinfo {volume} {59}},\ \bibinfo {pages} {8875}
  (\bibinfo {year} {1999})}\BibitemShut {NoStop}%
\bibitem [{\citenamefont {Dynes}\ \emph {et~al.}(1984)\citenamefont {Dynes},
  \citenamefont {Garno}, \citenamefont {Hertel},\ and\ \citenamefont
  {Orlando}}]{Dynes}%
  \BibitemOpen
  \bibfield  {author} {\bibinfo {author} {\bibfnamefont {R.~C.}\ \bibnamefont
  {Dynes}}, \bibinfo {author} {\bibfnamefont {J.~P.}\ \bibnamefont {Garno}},
  \bibinfo {author} {\bibfnamefont {G.~B.}\ \bibnamefont {Hertel}}, \ and\
  \bibinfo {author} {\bibfnamefont {T.~P.}\ \bibnamefont {Orlando}},\ }\href
  {\doibase 10.1103/PhysRevLett.53.2437} {\bibfield  {journal} {\bibinfo
  {journal} {Phys. Rev. Lett.}\ }\textbf {\bibinfo {volume} {53}},\ \bibinfo
  {pages} {2437} (\bibinfo {year} {1984})}\BibitemShut {NoStop}%
\bibitem [{\citenamefont {Pekola}\ \emph {et~al.}(2010)\citenamefont {Pekola},
  \citenamefont {Maisi}, \citenamefont {Kafanov}, \citenamefont {Chekurov},
  \citenamefont {Kemppinen}, \citenamefont {Pashkin}, \citenamefont {Saira},
  \citenamefont {M\"ott\"onen},\ and\ \citenamefont {Tsai}}]{PekolaDO}%
  \BibitemOpen
  \bibfield  {author} {\bibinfo {author} {\bibfnamefont {J.~P.}\ \bibnamefont
  {Pekola}}, \bibinfo {author} {\bibfnamefont {V.~F.}\ \bibnamefont {Maisi}},
  \bibinfo {author} {\bibfnamefont {S.}~\bibnamefont {Kafanov}}, \bibinfo
  {author} {\bibfnamefont {N.}~\bibnamefont {Chekurov}}, \bibinfo {author}
  {\bibfnamefont {A.}~\bibnamefont {Kemppinen}}, \bibinfo {author}
  {\bibfnamefont {Y.~A.}\ \bibnamefont {Pashkin}}, \bibinfo {author}
  {\bibfnamefont {O.-P.}\ \bibnamefont {Saira}}, \bibinfo {author}
  {\bibfnamefont {M.}~\bibnamefont {M\"ott\"onen}}, \ and\ \bibinfo {author}
  {\bibfnamefont {J.~S.}\ \bibnamefont {Tsai}},\ }\href {\doibase
  10.1103/PhysRevLett.105.026803} {\bibfield  {journal} {\bibinfo  {journal}
  {Phys. Rev. Lett.}\ }\textbf {\bibinfo {volume} {105}},\ \bibinfo {pages}
  {026803} (\bibinfo {year} {2010})}\BibitemShut {NoStop}%
\bibitem [{\citenamefont {Chockalingam}\ \emph {et~al.}(2009)\citenamefont
  {Chockalingam}, \citenamefont {Chand}, \citenamefont {Kamlapure},
  \citenamefont {Jesudasan}, \citenamefont {Mishra}, \citenamefont {Tripathi},\
  and\ \citenamefont {Raychaudhuri}}]{PhysRevB.79.094509}%
  \BibitemOpen
  \bibfield  {author} {\bibinfo {author} {\bibfnamefont {S.~P.}\ \bibnamefont
  {Chockalingam}}, \bibinfo {author} {\bibfnamefont {M.}~\bibnamefont {Chand}},
  \bibinfo {author} {\bibfnamefont {A.}~\bibnamefont {Kamlapure}}, \bibinfo
  {author} {\bibfnamefont {J.}~\bibnamefont {Jesudasan}}, \bibinfo {author}
  {\bibfnamefont {A.}~\bibnamefont {Mishra}}, \bibinfo {author} {\bibfnamefont
  {V.}~\bibnamefont {Tripathi}}, \ and\ \bibinfo {author} {\bibfnamefont
  {P.}~\bibnamefont {Raychaudhuri}},\ }\href {\doibase
  10.1103/PhysRevB.79.094509} {\bibfield  {journal} {\bibinfo  {journal} {Phys.
  Rev. B}\ }\textbf {\bibinfo {volume} {79}},\ \bibinfo {pages} {094509}
  (\bibinfo {year} {2009})}\BibitemShut {NoStop}%
\bibitem [{\citenamefont {Saira}\ \emph {et~al.}(2012)\citenamefont {Saira},
  \citenamefont {Kemppinen}, \citenamefont {Maisi},\ and\ \citenamefont
  {Pekola}}]{saira}%
  \BibitemOpen
  \bibfield  {author} {\bibinfo {author} {\bibfnamefont {O.-P.}\ \bibnamefont
  {Saira}}, \bibinfo {author} {\bibfnamefont {A.}~\bibnamefont {Kemppinen}},
  \bibinfo {author} {\bibfnamefont {V.~F.}\ \bibnamefont {Maisi}}, \ and\
  \bibinfo {author} {\bibfnamefont {J.~P.}\ \bibnamefont {Pekola}},\ }\href
  {\doibase 10.1103/PhysRevB.85.012504} {\bibfield  {journal} {\bibinfo
  {journal} {Phys. Rev. B}\ }\textbf {\bibinfo {volume} {85}},\ \bibinfo
  {pages} {012504} (\bibinfo {year} {2012})}\BibitemShut {NoStop}%
\bibitem [{\citenamefont {Wolf}(1985)}]{Wolf}%
  \BibitemOpen
  \bibfield  {author} {\bibinfo {author} {\bibfnamefont {E.~L.}\ \bibnamefont
  {Wolf}},\ }\href@noop {} {\emph {\bibinfo {title} {Principles of electron
  tunneling spectroscopy}}}\ (\bibinfo  {publisher} {Oxford University Press,
  Oxford},\ \bibinfo {year} {1985})\BibitemShut {NoStop}%
\bibitem [{\citenamefont {Mitrovi\'c}\ and\ \citenamefont
  {Rozema}(2008)}]{Mitrovic}%
  \BibitemOpen
  \bibfield  {author} {\bibinfo {author} {\bibfnamefont {B.}~\bibnamefont
  {Mitrovi\'c}}\ and\ \bibinfo {author} {\bibfnamefont {L.~A.}\ \bibnamefont
  {Rozema}},\ }\href {http://stacks.iop.org/0953-8984/20/i=1/a=015215}
  {\bibfield  {journal} {\bibinfo  {journal} {J. Phys.: Cond. Matter}\ }\textbf
  {\bibinfo {volume} {20}},\ \bibinfo {pages} {015215} (\bibinfo {year}
  {2008})}\BibitemShut {NoStop}%
\bibitem [{\citenamefont {Noguchi}\ \emph {et~al.}(2011)\citenamefont
  {Noguchi}, \citenamefont {Suzuki},\ and\ \citenamefont {Tamura}}]{Noguchi}%
  \BibitemOpen
  \bibfield  {author} {\bibinfo {author} {\bibfnamefont {T.}~\bibnamefont
  {Noguchi}}, \bibinfo {author} {\bibfnamefont {T.}~\bibnamefont {Suzuki}}, \
  and\ \bibinfo {author} {\bibfnamefont {T.}~\bibnamefont {Tamura}},\ }\href
  {\doibase 10.1109/TASC.2010.2089033} {\bibfield  {journal} {\bibinfo
  {journal} {IEEE Trans. Appl. Supercond.}\ }\textbf {\bibinfo {volume} {21}},\
  \bibinfo {pages} {756 } (\bibinfo {year} {2011})}\BibitemShut {NoStop}%
\bibitem [{\citenamefont {van Dover}\ \emph {et~al.}(1982)\citenamefont {van
  Dover}, \citenamefont {Bacon},\ and\ \citenamefont {Sinclair}}]{dover:764}%
  \BibitemOpen
  \bibfield  {author} {\bibinfo {author} {\bibfnamefont {R.~B.}\ \bibnamefont
  {van Dover}}, \bibinfo {author} {\bibfnamefont {D.~D.}\ \bibnamefont
  {Bacon}}, \ and\ \bibinfo {author} {\bibfnamefont {W.~R.}\ \bibnamefont
  {Sinclair}},\ }\href {\doibase 10.1063/1.93670} {\bibfield  {journal}
  {\bibinfo  {journal} {Appl. Phys. Lett.}\ }\textbf {\bibinfo {volume} {41}},\
  \bibinfo {pages} {764} (\bibinfo {year} {1982})}\BibitemShut {NoStop}%
\bibitem [{\citenamefont {O'Neil}\ \emph {et~al.}(2012)\citenamefont {O'Neil},
  \citenamefont {Lowell}, \citenamefont {Underwood},\ and\ \citenamefont
  {Ullom}}]{oneil}%
  \BibitemOpen
  \bibfield  {author} {\bibinfo {author} {\bibfnamefont {G.~C.}\ \bibnamefont
  {O'Neil}}, \bibinfo {author} {\bibfnamefont {P.~J.}\ \bibnamefont {Lowell}},
  \bibinfo {author} {\bibfnamefont {J.~M.}\ \bibnamefont {Underwood}}, \ and\
  \bibinfo {author} {\bibfnamefont {J.~N.}\ \bibnamefont {Ullom}},\ }\href
  {\doibase 10.1103/PhysRevB.85.134504} {\bibfield  {journal} {\bibinfo
  {journal} {Phys. Rev. B}\ }\textbf {\bibinfo {volume} {85}},\ \bibinfo
  {pages} {134504} (\bibinfo {year} {2012})}\BibitemShut {NoStop}%
\end{thebibliography}
\end{document}